\documentclass[pra,aps,amsfonts,amsmath,superscriptaddress,%
  twocolumn,showpacs,floatfix]{revtex4}
\usepackage{graphicx,color}
\usepackage{amsfonts}
\usepackage{amsmath}
\usepackage{bm}
\usepackage{epsfig}

\newcommand{\red}[1]{\textcolor[rgb]{1.00,0.00,0.00}{#1}}

\begin{document}
\title{Evolution of the proton $sd$
states in neutron-rich Ca isotopes}

\author{M. Grasso}
\affiliation{Institut de Physique Nucl\'eaire,
 Universit\'e Paris-Sud,
IN$_2$P$_3$-CNRS,
F-91406 Orsay Cedex, France}
\affiliation{Dipartimento di Fisica e Astronomia and INFN, Via Santa
  Sofia 64, I-95123 Catania, Italy}
\author{Z.Y. Ma}
\affiliation{China Center of Advanved Science and Technology (World
Laboratory), Beijing 100080, China}
\affiliation{China Institute of Atomic Energy, Beijing 102413, China}
\author{E. Khan}
\affiliation{Institut de Physique Nucl\'eaire, Universit\'e Paris-Sud,
IN$_2$P$_3$-CNRS, F-91406 Orsay Cedex, France}
\author{J. Margueron}
\affiliation{Institut de Physique Nucl\'eaire,
 Universit\'e Paris-Sud,
IN$_2$P$_3$-CNRS,
 F-91406 Orsay Cedex, France}
\author{N. Van Giai}
\affiliation{Institut de Physique Nucl\'eaire,
 Universit\'e Paris-Sud,
IN$_2$P$_3$-CNRS,
 F-91406 Orsay Cedex, France}
\begin{abstract}
We analyze the evolution with increasing isospin asymmetry of the
proton single-particle states $2s1/2$ and $1d3/2$ in Ca isotopes,
using non-relativistic and relativistic mean field approaches.
Both models give similar trends and it is shown that this
evolution is sensitive to the neutron shell structure, the two
states becoming more or less close depending on the neutron
orbitals which are filled. In the regions where the states get
closer some parametrizations predict an inversion between them. This
inversion occurs near $^{48}$Ca as well as very far from stability
where the two states systematically cross each other if the drip
line predicted in the model is located far enough. We study in
detail the modification of the two single-particle energies
 by using the equivalent potential in
the Schroedinger-like Skyrme-Hartree-Fock equations.
 The role played by central, kinetic and spin-orbit contributions
is discussed.
 We finally show that the effect of
a tensor component in the effective interaction
 considerably favors the inversion of the two proton states in $^{48}$Ca.
\end{abstract}
\maketitle

\section{Introduction}

Novel properties and new scenarios are expected for nuclei situated far from
stability.
The new generation of radioactive beam
facilities will allow to answer many open questions
about the pecularities of these unstable systems.
One of the major issues in the physics of exotic nuclei is
the study
of shell structure and magicity evolution when approaching the
drip lines \cite{dob,emilia}.
 From a theoretical
point of view, two aspects have been underlined as mainly responsible for the
evolution of single-particle energies
far from stability, the one-body 
spin-orbit potential
which is strongly modified when the surface becomes more diffuse 
 \cite{dob}
and the tensor force between neutrons and protons in valence 
sub-shells \cite{otsu1}. 

Recently, the $N=28$ shell closure has been experimentally analyzed
in the $^{46}$Ar($d,p$)$^{47}$Ar transfer reaction \cite{gaude}.
A strong reduction of the
neutron  $p$
 spin-orbit splitting
 has been observed in $^{47}$Ar with respect
to the isotone $^{49}$Ca.
Since $p$ states are
mainly localized in the interior of the nucleus, this
 strong reduction
cannot be justified by the presence of a diffuse surface which
would affect only high-$l$ states mainly concentrated at the
surface. A theoretical analysis based on the relativistic mean
field (RMF) approach has been proposed by Todd-Rutel et al.
\cite{topi1}. It predicts a strong reduction of the
spin-orbit splitting for neutron $2p$-states
in $^{46}$Ar as compared to $^{48}$Ca.
At $Z=20$, the  state $2s1/2$ is usually located between
$1d5/2$ and $1d3/2$.
 In the RMF calculations of Ref. \cite{topi1},
however, $2s1/2$ is predicted less bound
than $1d3/2$ in both $^{46}$Ar and $^{48}$Ca ($2s1/2-1d3/2$ inversion).
 In this scenario, $2s1/2$ is empty in $^{46}$Ar
and
occupied in $^{48}$Ca: thus, the
proton density profile in
$^{46}$Ar presents a strong depletion in the interior of the nucleus.
This reduction of the charge density in the center
would be responsible for
the modification of the spin-orbit in the nuclear interior and, \red{hence},
 for the
reduction of the neutron $2p$-splitting.

This problem of $2s1/2-1d3/2$ inversion of the proton states
has been already analyzed by
 Campi and
Sprung \cite{campi} within the Hartree-Fock (HF) + BCS model with an interaction derived from a G-matrix \cite{inte}.
$^{36}$Ar was found as a candidate for this inversion.
Skyrme forces do not
predict any inversion in this nucleus.
 It is thus worthwhile to revisit the problem for other nuclei in this region of the nuclear chart in the framework of the Skyrme-HF model.

In this work, we analyze the evolution of the
$s-d$ proton single-particle states in Ca
isotopes and the possible $2s1/2-1d3/2$ inversions.
We also present some comparisons with the corresponding results obtained
within RMF.
We neglect pairing in our treatment since Ca isotopes are proton
closed-shell nuclei.
We have checked that, within RMF 
the inclusion of neutron pairing does not
modify in a significant way the evolution of the proton states we are
interested in. The only important effect due to pairing is the shift of the
drip line towards heavier isotopes (for example, the drip line is shifted
from $^{60}Ca$ to $^{76}$Ca
with the parametrization NL3 \cite{nl3}). However,
this aspect is not
relevant in the present analysis which is not intended
to make any prediction on the
drip line position.
We choose the Ca isotopes since experimental signals for the inversion 
phenomena have
been found at least
in one of these isotopes,
$^{48}$Ca: the ground state  of $^{47}$K (one proton
less than $^{48}$Ca) is $1/2^+$ with a 
large spectroscopic factor
\cite{expk1} and
 the single-particle spectrum of $^{48}$Ca has been measured,
the proton state $1d3/2$ being more bound
than $2s1/2$ by about 300 keV  \cite{calcium}.
In our analysis, we explore all the contributions, kinetic, central,
spin-orbit and tensor,
which can modify the single-particle
energies with increasing $A$ and we show that not only the spin-orbit and tensor
terms are determinant.
The role of the central mean field term is in particular discussed.
Within the models which predict the crossing between the two states, we show
that this inversion occurs near $^{48}$Ca as well as in very neutron-rich
nuclei close to the drip line.

The article is organized as follows. In Sec. II we study the evolution
with increasing $A$ of the
difference $\Delta \epsilon$ between the
$2s1/2$ and $1d3/2$ energies obtained within
non-relativistic and relativistic approaches. In Sec. III  we
concentrate on the non-relativistic case and perform a detailed analysis of the
results. The different contributions to $\Delta \epsilon$ are
isolated by
analyzing them with the equivalent potential in the
Schroedinger-like HF equations.
In Sec. IV the effect of the tensor force
 is estimated in the framework of the SLy5-HF \cite{chaba} model.
Finally, conclusions are drawn in Sec. V.

\section{Evolution of \red{ $2s1/2$ and $1d3/2$ proton} states
within Skyrme-HF and RMF}

We first perform a
preliminary study with HF calculations of $^{48}$Ca using different
 Skyrme interactions. We
then choose three representative forces: SkI5 \cite{rei} which
predicts a $2s1/2-1d3/2$ inversion 
with an energy difference $\Delta \epsilon$
of $\sim$ 800 keV,
 SGII \cite{giai1}
which also reproduces the inversion ($\Delta \epsilon$ $\sim$ 200 keV)
and SLy4 \cite{chaba} for which there is no inversion.
With the three selected parametrizations we
have systematically analyzed the Ca isotopes
from $^{40}$Ca up to
the HF two-neutron drip line.
We recall that, in the three
considered Skyrme parametrizations there is no explicit tensor force.

We show in Fig. 1 the
difference $\Delta \epsilon$ between the energies of the proton states
$2s1/2$ and $1d3/2$ for the three Skyrme forces. The inversion
takes place where  $\Delta \epsilon$ is positive.
Corrections to the individual energies due to the coupling of single
particle motion with collective vibrations, which are neglected in our
treatment, should be expected (see, for instance, Ref. \cite{donati}).
However, by considering the energy difference instead of the
individual single-particle
energies the effects of these corrections should be reduced, the coupling to vibrations
having the tendency of shifting upwards the energies of both occupied states.

We observe that the SLy4-HF calculations never predict
the $2s1/2$-$1d3/2$ inversion. On the other hand, both
SkI5-HF and SGII-HF predict this inversion
around $^{48}$Ca as well as for more neutron-rich isotopes
starting from $^{58}$Ca up to the drip line.
The HF two-neutron drip line is located at $^{82}$Ca, $^{78}$Ca and
$^{60}$Ca with SkI5, SGII and SLy4, respectively.
The two experimental points for  $^{40}$Ca and  $^{48}$Ca are also
included in the figure.
The three sets of results globally present the same
behavior.
Indeed, in all three cases the 
quantity $\Delta \epsilon$ starts from a negative value and 
increases from A=40 to A=48. 
This generates a $2s1/2$-$1d3/2$ inversion 
with SkI5 (in $^{46}$Ca, $^{48}$Ca and $^{50}$Ca) and with SGII
 (in $^{48}$Ca).
 Going from $^{48}$Ca up to $^{52}$Ca the states cross again with SkI5 and
SGII, while the distance between them
increases with SLy4.
Beyond $^{52}$Ca $\Delta \epsilon$ increases again 
with the three
parametrizations. This generates another inversion
with SkI5 and SGII starting from $^{58}$Ca.
We notice also that
the nuclei for which $\Delta \epsilon$ presents maxima or minima are the same
for the three Skyrme forces.

\begin{figure}
\begin{center}
\epsfig{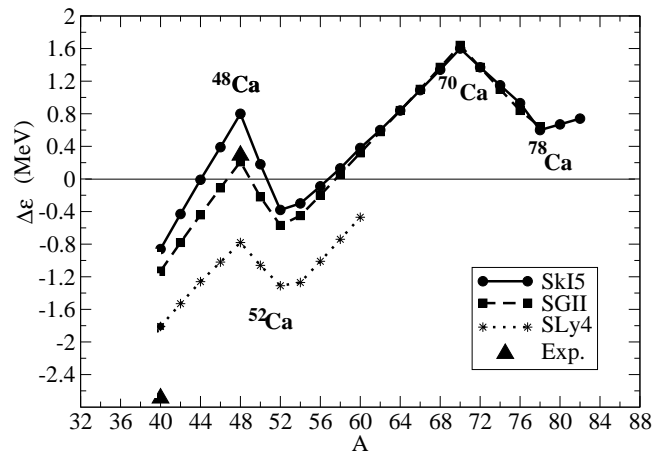}
\end{center}
\caption{Difference between the energies of the $2s1/2$ and $1d3/2$
proton states calculated with the
Skyrme interactions SkI5, SGII and SLy4 for Ca isotopes.}
\label{fig1}
\end{figure}

A natural question to ask is whether the above general trends are specific of the Skyrme-HF approach. It is well known that the RMF approach gives a spin-orbit potential whose (N-Z) dependence is somewhat different from that of Skyrme-HF models\cite{Ring-s.o.}.
We have performed RMF calculations with different parametrizations for
the same set of Ca isotopes using the
parametrizations DDME1 \cite{ddme1}, NL3 \cite{nl3} and NLB2 \cite{nlb2}. 
The latter one is chosen as an example of RMF model which does not lead to a 
$2s1/2$-$1d3/2$ level inversion. 
The calculated values of $\Delta \epsilon$ are shown
in Fig. 2 up to    
$^{60}$Ca which is the two-neutron drip line isotope predicted by DDME1 and 
NL3. 
Globally, we observe for
$\Delta \epsilon$ the same trend as that obtained within the non
relativistic HF, with maxima and minima corresponding to the same nuclei,
$^{48}$Ca and $^{52}$Ca.
Since comparable trends are obtained in both non-relativistic and relativistic approaches
we conclude that 
the calculated evolution of $2s1/2$ and $1d3/2$ states is a generic behavior.
We
can thus explore more in detail the results by
considering only the non-relativistic case.

\begin{figure}
\begin{center}
\epsfig{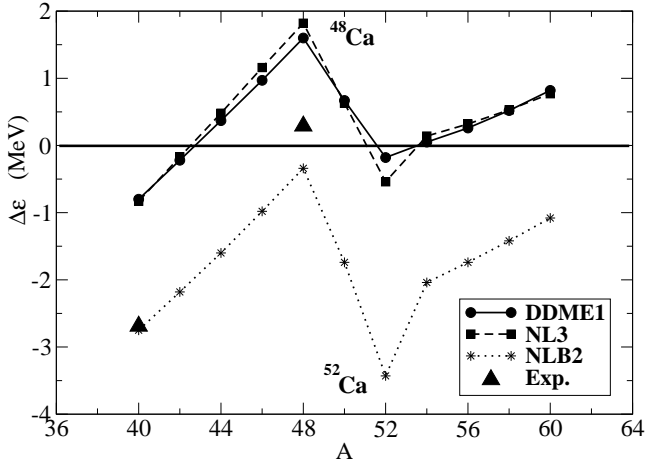}
\end{center}
\caption{Difference between the energies of the $2s1/2$ and $1d3/2$
proton states in Ca isotopes calculated in RMF with the parametrizations
DDME1, NL3 and NLB2.}
\label{fig2}
\end{figure}

\section{\red{Analysis of the contributions to $\Delta \epsilon$}}

We now concentrate on the maxima and minima of
$\Delta \epsilon$. 
 They correspond
to nuclei with neutron closed shells or sub-shells: 
the maximum at  $^{48}$Ca corresponds to the closure of the neutron
$1f7/2$ orbital whereas the minimum at $^{52}$Ca corresponds to the
filling of the neutron $2p3/2$ state. 
In the non relativistic Skyrme-HF model
the radial HF equations can be expressed in terms of an
energy-dependent equivalent potential $V_{eq}^{lj}$:
\begin{equation}
\frac{\hbar^2}{2m} \left[-\frac{d^2}{dr^2} \psi(r)
+\frac{l(l+1)}{r^2} \psi(r) \right] + V_{eq}^{lj} (r,\epsilon)
\psi(r)=\epsilon \psi(r)~, \label{new1}
\end{equation}
where 
\begin{equation}
V_{eq}^{lj}(r,\epsilon) = V_{eq}^{centr.} + \frac{m^*(r)}{m}
U_{so}^{lj}(r)
+ \left[1-\frac{m^*(r)}{m} \right] \epsilon~,
\label{new2}
\end{equation}
with $U_{so}^{lj}(r)=U_{so}(r) \times [j(j+1)-l(l+1)-3/4]$.
$U_{so}(r)$ is  the spin-orbit HF potential and $V_{eq}^{centr.}$ is
\begin{equation}
\begin{split}
V_{eq}^{centr.} =  \frac{m^*(r)}{m} U_0 (r)
-\frac{m^{*2}(r)}{2m\hbar^2} \left( \frac{\hbar^2}{2m^*(r)}
\right)^{'2}  \\ + \frac{m^*(r)}{2m} \left( \frac{\hbar^2}{2m^*(r)}
\right)^{''} ~,
\end{split}
\label{3}
\end{equation}
where $U_0(r)$ is the central
HF potentials and $m^*(r)$ is the effective
 mass \cite{chaba}. For protons $U_0$ includes the Coulomb
 potential. 
Up to a normalization  factor the HF radial wave function
$\phi$ of energy $\epsilon$ is related to the solution $\psi$
of Eq.(\ref{new1}) by the relation $\psi=(m^*/m)^{1/2}\phi$.
From Eqs. (\ref{new1})-
(\ref{3}) we can write $\Delta \epsilon$ as \begin{equation}
\begin{split}
\Delta \epsilon = \left[
\frac{\langle T \rangle_s}{\langle m^*/m \rangle_s}
-\frac{\langle T \rangle_d}{\langle m^*/m \rangle_d}  \right] \\
+\left[
\frac{ \langle V_{eq}^{centr.}\rangle_s }{\langle m^*/m \rangle_s}
-\frac{\langle V_{eq}^{centr.}\rangle_d}{\langle m^*/m \rangle_d}
\right]
\\
- \frac{\langle (m^*/m)
U_{so}^{d3/2} \rangle}{\langle m^*/m \rangle_d}
 ~,
\end{split}
\label{e5}
\end{equation}
where $T$ is the kinetic contribution. The 3 terms of the
r.h.s. 
of Eq. (\ref{e5}) - kinetic, central and spin-orbit - are plotted
in Fig. 3 for the force SkI5 and the nuclei $^{40}$Ca, $^{48}$Ca,
$^{52}$Ca and $^{70}$Ca. Similar results are obtained with SLy4 and
SGII. We mention that the mean value of the effective mass in the
denominators of Eq. (\ref{e5}) 
has very little A-dependence 
from $^{40}$Ca to$^{70}$Ca.
\begin{figure}
\begin{center}
\epsfig{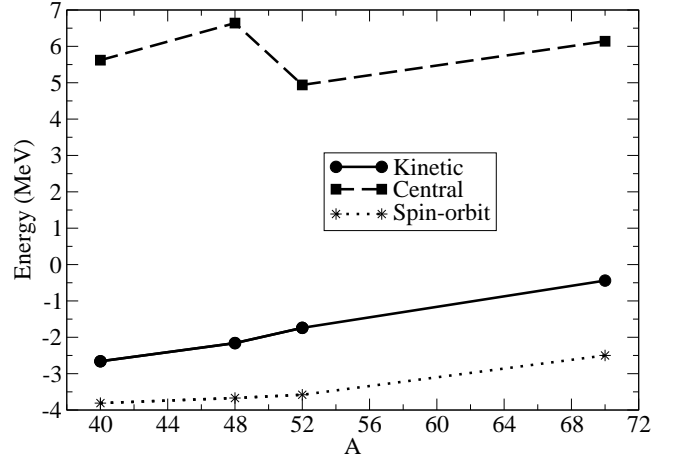}
\end{center}
\caption{Kinetic, central
and spin-orbit contributions of Eq. (\ref{e5}) for
SkI5 in  $^{40}$Ca,
$^{48}$Ca, $^{52}$Ca and $^{70}$Ca.}
\label{fig3}
\end{figure}
From Fig. 3, one notices that the spin-orbit and kinetic terms
 present a regular behavior as
a function of $A$.
Both of them are weakened with increasing isospin asymmetry
favouring the inversion in very neutron-rich isotopes. The
spin-orbit term is weakened because  
 the neutron surface becomes more diffuse with increasing $A$. In
general, the kinetic energy of an orbital depends on the mean
distance between its single-particle energy and the 
bottom of the potential in the region where the wave function
is localized. For the two 
states $2s1/2$ and $1d3/2$ we can look at 
the difference $\epsilon_{lj} - V_{eq}^{lj}(r_0)$, where 
$r_0$ is the root mean square (r.m.s.) radius of the corresponding
wave function.
 The evolution of $V_{eq}^{lj}$  with increasing $A$ is
governed by two effects: i) the lowering of the proton potential due
to the symmetry term; ii) the formation of a neutron skin which
modifies the proton distribution by pulling it towards larger
radii. The intensity of these two effects depends on the quantum
numbers of the neutron orbitals which are filled and of the proton
wave function under study. As an illustration, we consider $^{52}$Ca
and $^{70}$Ca. The r.m.s. radii $r_0$ and the values $\epsilon_{lj}
- V_{eq}^{lj}(r_0)$ are shown in Table I for the $2s1/2$ and
$1d3/2$ states. 
From $^{52}$Ca to $^{70}$Ca the difference $\epsilon_{lj} -
V_{eq}^{lj}(r_0)$ is reduced  more for $1d3/2$ (4.9$\%$) than for
$2s1/2$ (0.4$\%$). This analysis is confirmed by the evolution of
the two r.m.s. radii.
\begin{table}
\begin{center}
\begin{tabular}{|c|c|c|c|c|}
\hline
 $A$ & $r_0$ (fm)  & $r_0$ (fm)  &
$\epsilon -V_{eq}^{lj}(r_0)$ (MeV)   &
$\epsilon -V_{eq}^{lj}(r_0)$ (MeV)  \\
    &  $2s1/2$ & $1d3/2$ &  $2s1/2$  & $1d3/2$ \\
\hline
52 & 3.71 & 3.70 & 23.11 & 19.73 \\
\hline
70 & 3.81 & 3.89 & 23.01 & 18.76 \\
\hline
\end{tabular}
\end{center}
\caption{The values of  
$r_0$ and $\epsilon -V_{eq}^{lj}(r_0)$,  
for the states $2s1/2$ 
and $1d3/2$ 
in $^{52}$Ca and $^{70}$Ca. The interaction is SkI5.} \label{tab1}
\end{table}
It is evident that, going from $^{52}$Ca  to $^{70}$Ca, the $1d3/2$
wave function is more affected than  $2s1/2$ by the enlarging of the
potential due to the formation of a thick neutron skin. This
explains the increase of the kinetic contribution to
$\Delta\epsilon$ with the neutron number. We consider now the
central term of Eq.(\ref{e5}) which is responsible for the
maxima and minima of $\Delta \epsilon$, and
concentrate on the maximum at $^{48}$Ca. 
We mention that the major role played by the central term in 
modifying the single particle energies has been also  
underlined by 
Gaudefroy et al. \cite{gaude}.
We introduce the quantities $V_0$, $V_1$ and $V_2$ which
correspond to the contributions of the 3 terms of Eq.
(\ref{3}). 
They are plotted in Fig. 4 
for $^{40}$Ca, $^{48}$Ca and $^{52}$Ca. It turns out that the term
mainly affected by the neutron shell structure is $V_0$
which
\begin{figure}
\begin{center}
\epsfig{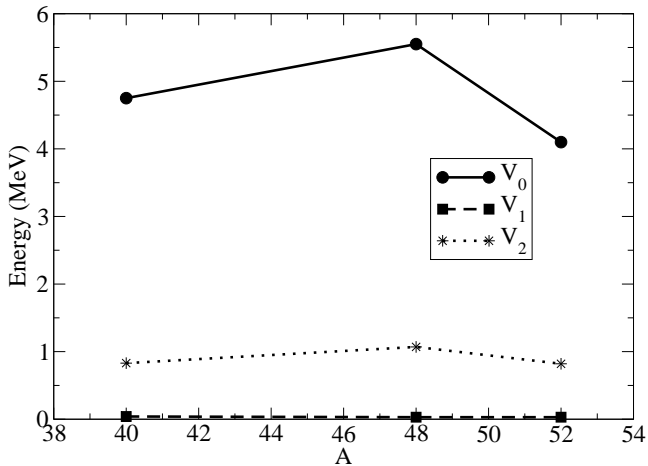}
\end{center}
\caption{$V_0$, $V_1$ and $V_2$ calculated with
SkI5 in  $^{40}$Ca,
$^{48}$Ca and $^{52}$Ca.}
\label{fig4}
\end{figure}
contains the Hartree-Fock potential.

We can separate the energy contributions of the N=Z=20 core
from those of the excess neutrons. For instance, the total nucleon
density $\rho$ is a sum of $\rho_{core}$ and $\rho_{excess}$, and
similarly for the other types of densities. Then, for any HF
quantity the core contribution is obtained by replacing in its
expression the total densities by core densities whereas the neutron
excess contribution corresponds to the rest. We show in Fig. 5 the
core and neutron excess contributions to  $V_0$.
\begin{figure}
\begin{center}
\epsfig{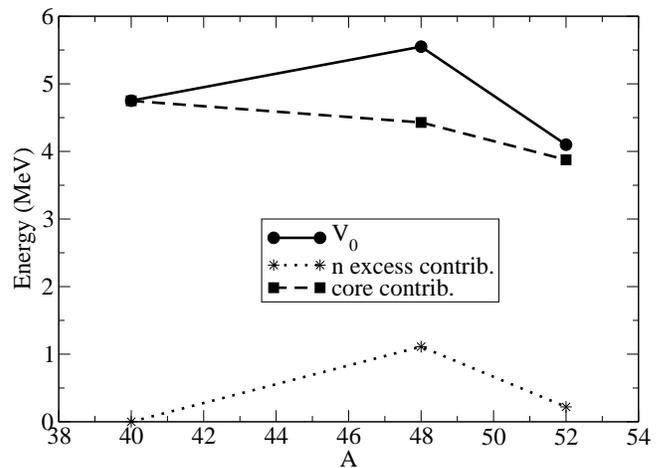}
\end{center}
\caption{Core and neutron excess contributions to $V_0$ for
SkI5 in  $^{40}$Ca,
$^{48}$Ca and $^{52}$Ca.}
\label{fig5}
\end{figure}
It is clear that the change of slope at $^{48}$Ca is mainly due to
the neutron excess contribution. We have further verified that the
term mainly responsible is the $t_0$ term of the Skyrme force. The
density-dependent term ($t_3$ term) is also sensitive to the neutron
shell structure but with an opposite behavior reducing the effect
due to the $t_0$ term. Hence, the main parameters which influence the
behavior of $\Delta\epsilon$  
are $x_0$ and $t_0$ as well as
$x_3$, $t_3$ and $\alpha$. We have checked that the role played by
the other terms of the HF potential is negligible.

To complete our analysis we consider separately the two
single-particle energies $2s1/2$ and $1d3/2$. We have verified
that the maximum of $\Delta \epsilon$ at $^{48}$Ca is mostly due to
the energy of the $1d3/2$ state which decreases less rapidly from
$^{48}$Ca to $^{52}$Ca than from $^{40}$Ca to $^{48}$Ca. This
behavior is explained in Fig. 6. In the top panels the neutron
excess contribution $P_{HF}$ to the HF potential due to the $t_0$
and $t_3$ terms is plotted for $^{44}$Ca, $^{48}$Ca and $^{52}$Ca
(see e.g. \cite{chaba} for the expressions in terms of the Skyrme
force parameters). In the middle panels we show
 the squares of the
 $1d3/2$ and $2s1/2$ radial
wave functions multiplied by $r^2$. In the bottom panels the
products $P_{HF}\vert \phi \vert ^2 r^2$ are shown.
Since $P_{HF}$ is negative (the $t_0$ contribution
 is negative and the largest in absolute value)
the proton single-particle energies are
 lowered with increasing N-Z, as expected.
We observe that, in $^{44}$Ca and $^{48}$Ca the potential related to
the neutron excess ($1f7/2$ neutron orbital) is concentrated in the
region where 
the $1d3/2$ wave function is localized. The overlap with this
wave function is thus the largest and this explains why the filling
of the neutron $1f7/2$ orbital has an important effect on the proton
$1d3/2$ energy which is strongly lowered. On the other hand, when
the neutron $2p3/2$ orbital is filled (from $^{48}$Ca to $^{52}$Ca)
the potential changes very little in the region where the
$1d3/2$ wave function is appreciable.
This explains why the
 $1d3/2$ energy decreases more from A=44 to 48 than from A=48 to 52.
The energy of the $2s1/2$ proton state is much less sensitive to the
neutron shell structure in these isotopes 
and it decreases rather monotonically from A=44 to 52.
\begin{figure}
\begin{center}
\epsfig{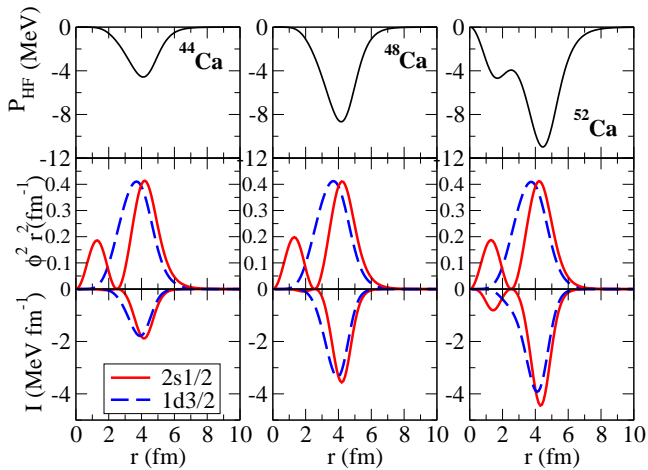}
\end{center}
\caption{Potential $P_{HF}$ (see text) (top), square of $1d3/2$ and
$2s1/2$ radial wave functions times $r^2$ (middle) and
their product 
 (bottom) calculated with
SkI5 in  $^{44}$Ca,
$^{48}$Ca and $^{52}$Ca.}
\label{fig6}
\end{figure}

\section{Tensor force effect}

The tensor force plays certainly a role in the evolution of single-particle states. This is discussed, e.g.,
in the framework of the shell model in ref. \cite{otsu1}. 
In a mean field approach, the tensor effect originates from the 
$\pi$-nucleon and $\rho$-nucleon contributions to the Fock terms 
\cite{giai,Long2007}, and it can be introduced phenomenologically in the 
parametrizations of effective interactions built for HF models 
\cite{Skyrme,stancu,otsu2}. Recent progress have been made in determining 
the tensor terms of Skyrme interactions \cite{colo,Brink2007} by adjusting 
the single-particle spectra measured in N=82 isotones and Z=50 isotopes 
\cite{schiffer}.

When this force is included the
spin-orbit potential presents an additional term depending on the
spin density $J$, namely,
\begin{equation}
U_{so}^q= \frac{W_0} {2} (\nabla \rho + \nabla \rho_q) +
\alpha J_q + \beta J_{q'} ~,
\end{equation}
where $q$ stands for neutrons (protons) and $q'$ for protons
(neutrons), $\alpha$ and $\beta$ consist of a sum of central and
tensor contributions: $\alpha=\alpha_C + \alpha_T$, $\beta=
\beta_C + \beta_T$. The central contributions depend only on the
velocity-dependent part of the Skyrme force whereas the tensor
contributions are generated by the tensor component of the Skyrme
force \cite{stancu,Brink2007}. 

To estimate the effect of the tensor force in our case we use the
Skyrme force SLy5 which already contains in the fitting protocol
the terms $\alpha_C$ and $\beta_C$.
 For $\alpha_T$ and $\beta_T$ we adopt the values determined in
Ref. \cite{colo} by comparing the Skyrme-HF predictions with
the data of Ref. \cite{schiffer}. These values are $\alpha_T =
-170$ MeV fm$^5$ and $\beta_T = 100$ MeV fm$^5$. We expect that
the tensor force favors the inversion in $^{48}$Ca (see for
instance Fig. 4 of Ref. \cite{otsu1}). Actually, from $^{40}$Ca to
 $^{48}$Ca the $1f7/2$ neutron orbital is filled. The interaction
between the proton orbital $1d3/2$ and the neutron orbital $1f7/2$
is attractive and its effect is to lower the energy of $1d3/2$,
thus favoring the crossing with $2s1/2$. As an illustration we
performed SLy5-HF calculations for  $^{40}$Ca, $^{48}$Ca,
$^{52}$Ca and  $^{70}$Ca ($^{70}$Ca is still bound within
SLy5-HF). We show in Fig 7 the values of $\Delta \epsilon$
obtained with and without the tensor contribution. As expected,
the tensor force increases the slope going from  $^{40}$Ca to
$^{48}$Ca bringing the two states close together 
and improving the agreement with the experimental data. The
improvement is quite strong since the value of $\Delta \epsilon$
in  $^{48}$Ca is equal to -0.26 MeV and -1.41 MeV with and without
the tensor contribution, respectively.

\begin{figure}
\begin{center}
\epsfig{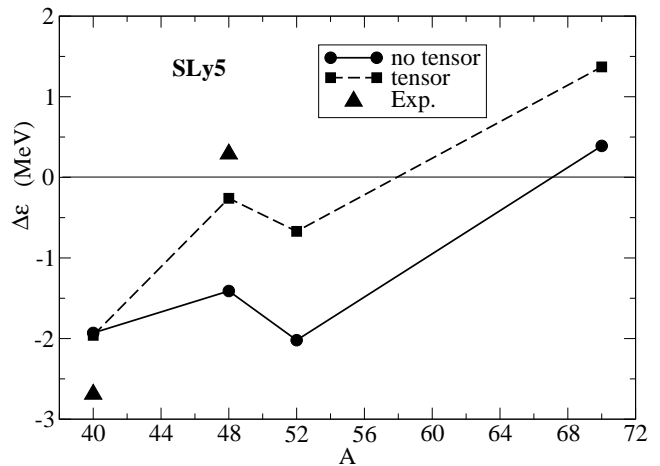}
\end{center}
\caption{Difference between the energies of $2s1/2$ and $1d3/2$
proton states in $^{40}$Ca,  $^{48}$Ca,  $^{52}$Ca and  $^{70}$Ca
calculated with SLy5 with and without the tensor contribution.}
\label{fig7}
\end{figure}

\section{Summary}
In this article we have analyzed the modification
of the proton single-particle states $2s1/2$ and $1d3/2$ in Ca
isotopes within the non-relativistic Skyrme-HF and the relativistic
RMF models. Pairing effects have been neglected since Ca isotopes
are proton closed-shell. We are interested in the evolution of
proton states and the inclusion of neutron pairing does not
affect significantly the global trend of our results. Both
models, HF and RMF, predict the same evolution  with increasing $A$
for the difference $\Delta \epsilon$ of the energies of the two
states. This evolution depends on the neutron orbitals which are
filled, $\Delta \epsilon$ presenting maxima and minima corresponding
to neutron shell or sub-shell closures. In particular, going
from $^{40}$Ca to $^{48}$Ca the two proton states come closer
to each other and they can sometimes cross in some models. By
performing an analysis based on the equivalent potential in the
non-relativistic Skyrme-HF approach, we have shown that
the kinetic and spin-orbit contributions present quite a regular
behavior with increasing $A$. They both strongly favor the inversion
of the two states in very neutron-rich nuclei. We have also
verified that the contribution which is mostly responsible for the
maximum of $\Delta\epsilon$ at $^{48}$Ca (and leading to
an inversion for some models) is the central HF potential and, in
particular the $t_0$ and $t_3$ terms. The former term favors the
crossing of the two states near $^{48}$Ca whereas the latter acts
against it. The net effect is that the two states get closer and
can cross each other in some models.

 We have finally
analyzed the role of the tensor force within the SLy5-HF model and
found that its contribution goes in the same direction as the $t_0$
term of the HF potential, favoring the inversion of the states near
$^{48}$Ca.

Our analysis was restricted to a purely mean field picture. Work
should be done to include effects beyond mean field. For
instance, particle-phonon coupling, which has been neglected here,
is expected to improve the quality of the theoretical predictions in
the study of single-particle states evolution.

\vspace{0.2cm}

The authors thank K. Bennaceur, A. Bhagwat, G. Col\`o, L. 
Gaudefroy, H. Sagawa and O. Sorlin 
for valuable discussions. ZYM and NVG acknowledge the partial
support of CNRS-IN2P3 (France) under the PICS program, of the
National Natural Science Foundation of China under Nos.10475116,
10535010 and the European Community project Asia-Europe Link in Nuclear 
Physics and Astrophysics CN/Asia-Link 008(94791).

\end{document}